\documentclass[reprint,floatfix,
showpacs,preprintnumbers,
nofootinbib,
nobibnotes,
amsmath,amssymb,aps,prd
]{revtex4-2}

\usepackage{multirow,hyperref}%
\usepackage[dvipsnames]{xcolor}%
\usepackage{multirow}%
\usepackage[final]{microtype}%
\usepackage{subcaption}%
\usepackage{amsmath,amssymb,amsfonts}%
\usepackage{amsthm}%
\usepackage{mathrsfs}%
\usepackage[title]{appendix}%
\usepackage{xcolor}%
\usepackage{textcomp}%
\usepackage{booktabs}%
\usepackage{algorithm}%
\usepackage{algorithmicx}%
\usepackage{algpseudocode}%
\usepackage{listings}%
\usepackage{graphicx}% Include figure files
\usepackage{dcolumn}% Align table columns on decimal point
\usepackage{bm}% bold math
\def\be{\begin{equation}}
\def\ee{\end{equation}}
\def\beq{\begin{eqnarray}}
\def\eeq{\end{eqnarray}}
\def\beqs{\begin{subequations}}
\def\eeqs{\end{subequations}}

\def\pa{\partial_a}
\def\pb{\partial_b}

\def\na{\nabla_a}
\def\nb{\nabla_b}

\def\T{\mathcal{T}}
\def\R{\mathcal{R}}
\def\A{\mathcal{A}}
\def\B{\mathcal{B}}
\def\Bi{(\mathcal{B}^{-1})}

\def\n{\nonumber}
\def\Tm{T_{\text{\tiny m}}}
\def\SLB{S_{\text{\tiny LBG}}}
\def\LLB{L_{\text{\tiny LBG}}}
\def\fLB{f_{\text{\tiny LBG}}}
\begin{document}

%%%%%%%%%%%%%%%%%%%%%%  preprint-version %%%%%%%%%%%%%%%%%%%

%\preprint{V3 -- 11/27/2024}

%%%%%%%%%%%%%%%%%%  title   %%%%%%%%%%%%%%%%%%%%%%%%%%%%%%

\title{Mimetic gravity in the extended objects framework}% Force line breaks with \\
%\thanks{A footnote of the article title}%

%%%%%%%%%%%%%%%%% authors %%%%%%%%%%%%%%%%%%%%%%%%%%%%%

\author{Efra\'\i n Rojas}
\email{efrojas@uv.mx}

%\author{G. Cruz}
%\email{giocruz@uv.mx}

\affiliation{Facultad de F\'\i sica, Universidad Veracruzana, 
Paseo No. 112, Desarrollo Habitacional Nuevo Xalapa, Xalapa-Enr\'\i quez, 91097, Veracruz, M\'exico
}%


%\affiliation{
%\affiliation{
%\affiliation{%


\date{\today}% It is always \today, today,

%%%%%%%%%%%%%%%%%%%%%  abstract  %%%%%%%%%%%%%%%%%%%%%%%%%%%%

\begin{abstract}
Starting from the most general second-order in derivatives 
theories describing extended objects of arbitrary dimension 
evolving geodetically in a codimension-one flat ambient 
space-time, we determine the subset of models yielding 
second-order equations of motion, forming an intriguing 
theory known as Lovelock-type brane gravity (LBG). These 
models further extend the so-called geodetic brane gravity 
(GBG) approach, thereby naturally promoting the GBG geometric 
properties, allowing LBG to be reformulated as a mimetic 
embedding gravity and, in turn, the possibility of 
introducing fictional matter through a peculiar current 
$\T^{a\,\mu}$. Grounded in the elasticity theory, we provide
a possible origin of such a current. Finally, variational 
techniques are employed to elucidate the mechanical function 
of both the dark current $\T^{a\,\mu}$ and its tangential 
components $\T^{ab}$; these serve as the constituents of 
a fictional energy-momentum tensor that shares 
characteristics with a perfect fluid.
%\begin{description}
%\end{description}
\end{abstract}

%%%%%%%%%%%%%%%%%%%%%%%   pacs   %%%%%%%%%%%%%%%%%%%%%%%%%%%%%%

%\pacs{02.30.Jr, 04.20.Fy, 11.10.Ef}% PACS, the Physics and Astronomy
\maketitle

%\tableofcontents

%%%%%%%%%%%%%%%%%%%%%%%%%%%%%%%%%%%%%%%%%%%%%%%%%%%%%%%%%%%%%%%%%%
\section{Introduction}
\label{sec:intro}
%%%%%%%%%%%%%%%%%%%%%%%%%%%%%%%%%%%%%%%%%%%%%%%%%%%%%%%%%%%%%%%%%%%

The problem of uncovering the nature of dark matter has been 
pursued by many authors through various approaches. Mimetic 
gravity is a diffeomorphism-invariant extension of General 
Relativity (GR) designed to emulate dark matter on large 
scales, through an emergent effect of the geometry of 
the space-time by changing the independent variables, without 
the need to introduce matter fields~\cite{Mukhanov2013}. 
Building on this idea, modified gravitational theories have 
been formulated, as controlled deviations from GR for this 
purpose, as evidenced by the so-called \textit{Geodetic 
Brane Gravity} (GBG) theory~\cite{Davidson2003}. GBG, also 
known as \textit{Embedding Gravity Theory}, based on the 
Regge-Teitelboim (RT) model is an interesting geometrical continuation of string theory that views our Universe as an 
extended object isometrically embedded, and geodesically 
floating, in a $N$-dimensional flat Minkowski space time, 
where the embedding vector $X^\mu$ acts as field variable 
rather than the metric tensor $g_{ab}$,~\cite{RT1975}. 
According to the isometric immersion theorems, at most 
$N = \frac{1}{2} n (n+1)$ is needed to locally embed a 
general $n$-dimensional metric, a number that can be reduced 
if the $n$ space-time admits some Killing vector fields, as 
in the case of embedding an FRW-type geometry~\cite{Kasner1921,Friedman1961,Rosen1965}. On physical grounds, 
GBG  has generated interest in research as an engaging 
formulation for describing gravity, its cosmological 
implications in extra dimensions, as a suitable alternative 
for the quantization of gravity~\cite{Deser1976, Pavsic1985,Tapia1989,Davidson2003,Davidson1998,Paston2007, Rojas2009,Estabrook2010,Banerjee2014,Rojas2022,Fabi2022,Stern2023}, 
and, since it is parametrized by a conserved bulk energy by a 
constant, $\omega$, it allows the introduction of additional 
fictitious matter, also called as \textit{dark matter} or 
\textit{embedding matter}, which seems to be a mere manifestation 
of the additional extra dimensions~\cite{Davidson2001,Paston2023,Stern2023,Rojas2024,Paston2026}.

Technically, GBG consists of an integral of the Ricci scalar, 
$\R$, over the world volume swept out by an extended object 
in the course of its evolution which may include matter content 
on the world volume through a Lagrangian $L_{\text{\tiny m}}$. 
Within the extended objects framework, the Ricci scalar 
depends linearly on the second-order derivatives of 
the embedding functions $X^\mu$, yielding, in common with the 
Dirac-Nambu-Goto theory,  second-order equations of motion 
(eom). This is the result of being an affine in acceleration 
theory~\cite{Rojas2016}. The GBG field equations are
\be 
\na \left[ \left( G^{ab} - \kappa T_{\text{\tiny m}}^{ab} 
\right)\pb X^\mu \right] = 0,
\label{eom0}
\ee
which admit extra solutions beyond standard GR. As discussed 
in~\cite{Pavsic1985,Davidson2003}, a more generalized solution is
\be 
\left( G^{ab} - \kappa T^{ab} \right)\pb X^\mu =
C^{a\,\mu},
\label{general}
\ee
where $C^{a\,\mu} = C^{ab} \pb X^\mu$ must satisfy 
$\na C^{a\,\mu} = 0$ implying that $\na C^{ab} = 0$ 
with $C^{ab} \neq 0$. Clearly, equations~(\ref{eom0}) are 
of second order in derivatives of $X^\mu$ and their form 
is unaltered by the presence of $C^{a\,\mu}$. By nature, 
$C^{ab}$ acts as the covariantly conserved energy-momentum 
tensor associated with a pressureless fluid~\cite{Pavsic1985,Paston2018}. 
It is natural to ask a simple-minded question. Do exist 
geometrical models for extended objects that depend on 
fundamental forms $g_{ab}$ and $K_{ab}$ whose equations 
of motion can be expressed in this way? 
Along these lines, it is worth investigating whether more 
general geometric models with these features can be found, 
and whether they can inherently involve such a fictitious 
form of dark matter. 
 
Relativistic extended objects of arbitrary dimension, 
are generalizations of particles and strings which 
attempt to represent many physical systems of an appropriate 
dimension, in terms of fields confined to their trajectories, 
propagating in a fixed background~\cite{Carter1992,defo1995}. 
In a geodetic setting, with no matter included, a brane can be 
slack and may wiggle and move, but its world volume will take 
on a certain shape, so the only relevant degrees of freedom (dof) 
should be those associated with its geometric configuration 
depending on how the world volume is embedded within the 
ambient spacetime. 
This fact leads to analyze its behaviour through Lagrangians 
constructed locally from the geometry of the world volume 
through geometrical invariants, composed from the fundamental 
forms associated to this surface, $L (g_{ab}, K_{ab})$. 
The presence of the extrinsic curvature signals the 
existence of second-order temporal and spatial derivatives 
of the field variables, $X^\mu$. This produces reluctance due 
to the emergence of non-physical dof that arise as a result of handling usual fourth-order equations of motion (eom) and 
therefore dealing with an unexpected number of dof. The RT 
model, being an affine in acceleration theory, falls 
into this category of gravity theories, but plays a key role 
in the understanding of richer geometric models leading to 
second-order eom.

In this paper, our purpose is twofold. Initially, we derive 
the form of the most general geometric second-order 
Lagrangian functions, which depend on the two fundamental 
forms, and which generate second-order eom. The resulting 
framework has been termed Lovelock-type brane gravity 
(LBG) and has been developed from a technical perspective, 
while its cosmological implications have been explored in some 
brane-world scenarios~\cite{Rojas2013,Rojas2016,Rojas2019,Rojas2025,Rojas2025b}. 
The eom are expressed in terms of conserved tensors, 
$J^{ab}_{(s)}$, which form the backbone of the linear 
momentum density of the extended object, and in such a way 
that makes it possible to introduce an extra term that leaves unaltered their form, through a geometric current $\T^{a\,\mu}$ 
on the world volume, and which entails the presence of a 
peculiar energy-momentum tensor. Second, we offer a possible 
explanation for the physical origin of this unusual current 
through the concept of internal forces, reminiscent of what 
occurs in non-relativistic mechanics, where the movement of 
the center of mass of the system remains unaltered. Finally, 
to strengthen this description, and following Paston~\cite{Paston2018}, we describe that LBG theory, as 
an extended version of GBG theory, allows for a variational 
formulation that incorporates the unusual current through 
Lagrange multipliers,  with their corresponding eom enforcing 
as the constraints the definition of the tangent vectors in 
terms of certain matrix structures, as well as the definition 
of the induced metric, thus ensuring that the dynamics 
remain unaltered.

LBG is a theory free from many of the pathologies that plague 
higher-order derivative theories thus ensuring no propagation 
of extra dof. The trade-off is that the resulting equation of 
motion is highly non-linear in $X^\mu$. The theory has led 
to interest in having potential physical applications, mainly 
at the cosmological level at late times, since it allows for  
considering alternative purely geometrical theories that might 
underlie the current puzzle of cosmic 
acceleration~\cite{Rojas2012,Rojas2024,Rojas2025b}. Although 
LBG is validated for codimension-1 scenarios, mostly to allow 
potential odd-degree extrinsic curvature polynomials that may 
modify the dynamics of the brane, when tailoring each model to specific geometries, one must not overlook the isometric 
embedding theorem to ensure the existence of a local embedding 
frame. In particular, a 5-dimensional Minkowski bulk is enough to 
host a 4-dimensional FRW brane,~\cite{Kasner1921,Friedman1961,Rosen1965}. 

This paper is organized as follows. In Sec.~\ref{sec2}, in 
order to make the paper self-contained, we quickly recap the geometrical structures defined on the brane trajectory and 
obtain the equations of motion arising from second-order 
Lagrangians of the form $L(g_{ab},K_{ab})$. In Sec.~\ref{sec3}, 
we derive all possible second-order Lagrangians $L(g_{ab},K_{ab})$ leading to second-order eom. A conjecture about the nature of 
the fictitious matter emerging from this geometrical scheme 
is examined in Sec.~\ref{sec4}. In Sec.~\ref{sec5}, the 
additional current and its energy-momentum tensor are 
incorporated into extended actions as Lagrange multipliers, 
yielding the complete set of equations of motion in accordance 
with the mimetic gravity view. Finally, in Sec.~\ref{sec6} we conclude with some remarks. An appendix is devoted to provide 
cumbersome variations needed in Sec.~\ref{sec5}.

%%%%%%%%%%%%%%%%%%%%%%%%%%%%%%%%%%%%%%%%%%%%%%%%%%%%%%%%
\section{Extended objects mechanics}
\label{sec2}
%%%%%%%%%%%%%%%%%%%%%%%%%%%%%%%%%%%%%%%%%%%%%%%%%%%%%%%

Consider a $p$-dimensional space-like brane, $\Sigma$, 
floating in a flat Minkowski spacetime, $\mathcal{M}$, 
of dimension $N=p+2$ with metric $\eta_{\mu\nu}$, \,\,($\mu,
\nu = 0,1,2,\ldots, p+1$). We set $x^\mu = X^\mu (x^a)$, 
with $x^\mu$ being local coordinates in $\mathcal{M}$, 
and the embedding functions $X^\mu$ specify the trajectory 
of $\Sigma$, that is, the world volume $m$, parametrized 
by the local coordinates $x^a$, \,\,$(a,b = 0,1,2, 
\ldots,p)$. $m$ is an oriented $(p+1)$-dimensional 
time-like  manifold. 

To describe the curved features of $m$ it is necessary to
derive them from $X^\mu$ with tensor calculus. Indeed,
the tangent space to $m$ is spanned by the $p+1$ tangent 
vectors $e^\mu{}_a := \partial_a X^\mu$ while the normal 
space is one-dimensional and is spanned by a single space-like 
vector $n^\mu$. This orthonormal basis is defined through 
the relations $e_a \cdot n = \eta_{\mu\nu} e^\mu{}_a n^\nu 
= 0$ and $n\cdot n = \eta_{\mu\nu} n^\mu n^\nu = 1$. 
Hereafter, a central dot will denote contraction with 
$\eta_{\mu\nu}$.
 
The basis $\{ e_a, n \}$ induces the induced metric and the 
extrinsic curvature, also named first and second fundamental
forms of $m$,~\cite{Spivak1970,defo1995}
\be 
\begin{aligned}
g_{ab} &= \pa X \cdot \pb X,
\\
K_{ab} &= - n \cdot \pa \pb X ,
\end{aligned}
\label{FSff}
\ee 
respectively. These structures characterize any surface since 
they encode the geometrically significant derivatives of 
$X^\mu$ whereby all world volume invariants can be generated 
from $g_{ab}$ and $K_{ab}$.

The dynamical behavior of $\Sigma$ will be described by an
action functional of the form
\be 
S[X^\mu] = \int_m d^{p+1}x\, \sqrt{-g} \,L(g_{ab}, K_{ab}),
\label{action}
\ee
where $g:= \det (g_{ab})$. It is assumed that the 
action~(\ref{action}) be invariant under reparametrizations of 
$m$ where $X^\mu$ are considered as the field variables of 
the theory~\cite{Capo2000}. Each effective geometrical model is  
a scalar built from the fundamental forms~(\ref{FSff}) and 
their covariant derivatives. Clearly, theories of the form~(\ref{action}) depend on second order derivatives of 
$X^\mu$ which commonly lead to fourth order eom for the field variables. Remarkably, under specific constraints, a number 
of these models yield second-order equations of motion, as we 
shall see shortly.

On using $\delta \sqrt{-g} = (\sqrt{-g}/2)g^{ab} \delta 
g_{ab}$, the variation of the action~(\ref{action}) reads
\be
\delta S = \int_m \sqrt{-g} \left[
\left( \frac{1}{2} L g^{ab} - H^{ab} \right) \delta g_{ab}
+ L^{ab} \delta K_{ab}
\right],
\label{var1}
\ee
where we have absorbed the differential $d^{p+1}x$ in the
integral and we have introduced the symmetric second rank 
tensors
\be 
H_{ab} := \frac{\partial L}{\partial g^{ab}},
\quad \text{and} \quad
L^{ab} := \frac{\partial L}{\partial K_{ab}}.
\label{tensors}
\ee
Under infinitesimal deformations $X^\mu \to X^\mu + \delta
X^\mu$ for $m$, to first order, the fundamental
forms change according to
\be 
\begin{aligned}
\delta g_{ab} &= 2 \partial_{(a} X \cdot \nabla_{b)} \delta X,
\\ 
\delta K_{ab} &= - n \cdot \na \nb \delta X,
\end{aligned}
\ee
where $\na$ is the world volume covariant derivative compatible
with $g_{ab}$. In terms of these, variation~(\ref{var1}) becomes
\beq 
\delta S &=& - \int_m \sqrt{-g} \na 
\left[( L g^{ab} - 2 H^{ab}) \pb X + \nb (L^{ab}\,n)
\right] 
\cdot \delta X
\n
\\
&+&
\int_m \sqrt{-g}\,\na Q^a,
\label{var2}
\eeq
where $Q^{a}$ incorporates all gathered terms in a divergence 
after integrations by parts. The symmetric tensors~(\ref{tensors}) fulfil the identity~\cite{Capo2000}
\be 
\label{id1}
2 H^{ab} = K^a{}_c L^{bc} + K^b{}_c L^{ac}.
\ee
Plugging this, and the Gauss-Weingarten (GW) equation $\na 
n^\mu = K_a{}^b \,\pb X^\mu$,~\cite{Spivak1970,defo1995}, 
into~(\ref{var2}), the variation can be cast in the form
\be 
\delta S = - \int_m \sqrt{-g} \, \na f^{a} \cdot \delta X
+ \int_m \sqrt{-g}\, \na Q^a,
\label{var3}
\ee
where we have introduced
\be 
\label{famu}
f^{a\,\mu} := 
\left( L\,g^{ab} - K^a{}_c L^{bc} \right)\pb X^\mu
+ (\nb L^{ab})\,n^\mu,
\ee
while $Q^a$ appearing in the total divergence takes the form
\be 
Q^a:= f^a \cdot \delta X - (L^{ab}\,n) \cdot \nb \delta X.
\label{Qa}
\ee
In the extended objects framework evolving in a flat Minkowski
space time, $f^{a\,\mu}$ is nothing but the stress conserved 
tensor for $m$. On physical grounds, as a result of the Noether theorem, this is related to the linear momentum associated with the extended object~\cite{Capo2000}, whereas from the elasticity theory viewpoint, this quantity is related to the normal projections of forces, per unit area, on the extended object generated by deformations~\cite{Landau1986}.

From~(\ref{var3}) we can read off the field equations through 
the conservation of $f^{a\,\mu}$,
\be 
\na f^{a\,\mu} = 0.
\label{eom1}
\ee
For reference later on it will be convenient to dismantle
this world volume divergence. On account of the GW equations 
$\na e^\mu{}_b = - K_{ab} n^\mu$ and $\na n^\mu = K_a{}^b\,
e^\mu{}_b$~\cite{Spivak1970,defo1995}, from~(\ref{famu})
and~(\ref{eom1}) we find
\beq 
\left[
\na \left( L\,g^{ab} - K^a{}_c L^{bc}\right) + \na L^{ac}\,
K_c{}^b 
\right] e^\mu{}_b
&+& 
\n
\\
\left[ 
\na \nb L^{ab} - (L\,g^{ab} - K^a{}_c L^{bc})K_{ab} 
\right]
n^\mu &=& 0.
\n
\eeq
Thus, we have decomposed the divergence-free property of 
$f^{a\,\mu}$ into a normal component, which 
acts as the physical equation of motion, and tangential 
components that capture the inherent gauge redundancy of 
the theory
\beq 
\na \nb L^{ab} - (L\,g^{ab} - K^a{}_c L^{bc})K_{ab} &=& 0,
\label{eom2}
\\
\na 
\left( L\,g^{ab} - K^a{}_c L^{bc}\right) + \na L^{ac}\,
K_c{}^b &=& 0,
\label{id2}
\eeq
respectively. One can generically assure from~(\ref{eom2})
that the solely equation of motion is fourth order in 
derivatives of the fields of interest, but this expression 
offers deeper geometric insight.  Regarding~(\ref{id2}), 
these can be considered as geometric identities, also named 
as Bianchi identities, satisfied by the fundamental forms.

%%%%%%%%%%%%%%%%%%%%%%%%%%%%%%%%%%%%%%%%%%%%%%%%%%%%%%%%
\subsection{Inclusion of matter}
\label{subsec2}
%%%%%%%%%%%%%%%%%%%%%%%%%%%%%%%%%%%%%%%%%%%%%%%%%%%%%%%%

In the study of brane cosmological scenarios, the inclusion 
of some matter content is required. If so, an action matter 
should be included in our description $S_{\text{\tiny m}} = 
\int_m \sqrt{-g} L_{\text{\tiny m}}$, with a matter Lagrangian $L_{\text{\tiny m}} (\varphi (x^a), X^\mu)$ where $\varphi 
(x^a)$ denotes matter fields living on the brane. The form of~(\ref{eom1}) remains practically unchanged since it only 
receives an extra contribution. A variational process 
applied to $S_{\textit{\tiny m}}$ yields $\delta 
S_{\textit{\tiny m}} = \int_m \left[ \partial (\sqrt{-g} 
L_{\text{\tiny m}})/ \partial g^{ab} \right] \delta 
g^{ab}$. After adding this to the variation~(\ref{var1}) 
followed by insertion of $\delta g^{ab} = - g^{ac}g^{bd} 
\delta g_{cd}$, we find the Euler-Lagrange 
derivative~(\ref{eom1}) with
\be 
f^{a\,\mu} = \left( L\,g^{ab} - K^a{}_c L^{bc} + 
\Tm^{ab} \right)\pb X^\mu + (\nb L^{ab})\,n^\mu,
\label{famu2}
\ee
where $T_{ab}^{\text{\tiny m}} = - (2/\sqrt{-g}) \partial
(\sqrt{-g} L_{\text{\tiny m}})/\partial g^{ab}$ is the
world volume energy-momentum tensor. Both the equation of motion~(\ref{eom2}), and the Bianchi identity~(\ref{id2}) 
result in
\beq 
\na \nb L^{ab} - \left[ (L\,g^{ab} - K^a{}_c L^{bc})
+ \Tm^{ab} \right] K_{ab} &=& 0,
\label{eom3}
\\
\na \left( L\,g^{ab} - K^a{}_c L^{bc} + \Tm^{ab} \right) 
+ \na L^{ac}\,K_c{}^b 
&=& 0,
\label{id3}
\eeq
respectively. A deeper dive into these structures follows.

%%%%%%%%%%%%%%%%%%%%%%%%%%%%%%%%%%%%%%%%%%%%%%%%%%%%%%%%%%%%%%%%
\section{Brane Lagrangians for second-order equations of motion}
\label{sec3}
%%%%%%%%%%%%%%%%%%%%%%%%%%%%%%%%%%%%%%%%%%%%%%%%%%%%%%%%%%%%%%%%

We will now focus on determining the general brane 
Lagrangians that yield second-order eom. In light 
of~(\ref{eom2}), it is readily inferred that the matrix 
$L^{ab}$ must either vanish identically or satisfy a 
special property. Given $L=L(g_{ab},K_{ab})$, it is 
expected that $L^{ab}$ inherits the dependence on 
second-order derivatives through $K_{ab}$. From~(\ref{eom2}), 
we readily note that to achieve this aim, $L^{ab}$ must 
be divergence-free
\be 
\na L^{ab} = 0.
\label{id4}
\ee
Therefore, the stress tensor~(\ref{famu}) becomes entirely tangential, which is the hallmark of physically consistent 
brane theories that yield second-order equations of motion. 
Condition~(\ref{id4}) entails that the dynamics of the 
extended object, given by~(\ref{eom2}) and~(\ref{id2}), 
must satisfy
\beq 
\left( L\,g^{ab} - K^a{}_c \frac{\partial L}{\partial 
K_{bc}} \right) K_{ab} &=& 0,
\label{eom4}
\\ 
\na \left( L\,g^{ab} - K^a{}_c \frac{\partial L}{\partial 
K_{bc}} \right) &=& 0.
\label{id5}
\eeq

According to matrix theory, given an $n \times n$ matrix, 
say $A^a{}_b$, the so-called $s$th \textit{discriminant} of 
$A^a{}_b$,~\cite{Tapia2007}, that is, the sum $A_{(s)}$ of 
all $(s\times s)$ principal minors of $\det (A^a{}_b)$, is 
defined by~\cite{Lovelock1989}
\be 
s! A_{(s)} = \delta^{a_1 a_2 \cdots a_s}_{b_1 b_2 \cdots b_s}
A^{b_1}{}_{a_1} A^{b_2}{}_{a_2} \cdots A^{b_s}{}_{a_s},
\label{discriminant}
\ee
with $s=1,2, \ldots, n.$, where we have employed the 
alternating tensor named generalized Kronecker delta (gKd), 
\be 
\delta^{a_1 a_2 a_3 \cdots a_s}_{b_1 b_2 b_3 \cdots b_s}
: =
\left|
\begin{matrix}
\delta^{a_1}{}_{b_1} & \delta^{a_1}{}_{b_2}
& \delta^{a_1}{}_{b_3} & \cdots &
\delta^{a_1}{}_{b_s}
\cr
\delta^{a_2}{}_{b_1} & \delta^{a_2}{}_{b_2}
& \delta^{a_2}{}_{b_3} & \cdots &
\delta^{a_2}{}_{b_s}
\cr
\vdots & \vdots & \ddots & \vdots & \vdots
\cr
\delta^{a_{s-1}}{}_{b_1} & \delta^{a_{s-1}}{}_{b_2}
& \delta^{a_{s-1}}{}_{b_3} & \cdots &
\delta^{a_{s-1}}{}_{b_s}
\cr
\delta^{a_{s}}{}_{b_1} & \delta^{a_{s}}{}_{b_2}
& \delta^{a_{s}}{}_{b_3} & \cdots &
\delta^{a_{s}}{}_{b_s}
\end{matrix}
\right|.
\ee
The following recurrence relation is straightforwardly
shown for the $s$th $A_{(s)}$ function
\be 
A_{(s)} \,g^{ab} - A^a{}_c \frac{\partial A_{(s)}}{\partial 
A_{bc}} = \frac{\partial A_{(s+1)}}{\partial A_{ab}}.
\label{id6}
\ee
The discriminants $A_{(s)}$ likewise fulfil the following 
property. For an $(n\times n)$ matrix, only the first $n$ discriminants are non-trivial, while those of an order higher 
than $n$ are identically zero, that is, $A_{(s)} = 0$ for 
$s>n$,~\cite{Tapia2007}.

The results of this matrix development serve our primary 
aim. It is clear that the fulfillment of~(\ref{eom4})
and~(\ref{id5}) is achieved if $L$ is related to the $s$th
discriminant of the matrix $K^a{}_b$, so that under the
identification $L \rightarrow L_s$ we obtain
\be  
\begin{aligned}
L^{ab}_{s+1} K_{ab} & = 0,
\\
\na L^{ab}_{s+1} & = 0.
\end{aligned}
\ee 
The conserved stress tensor~(\ref{famu}) reduces to
\be 
f^{a\,\mu}_{(s)} = L^{ab}_{s+1}\,\pb X^\mu.
\label{famus}
\ee 
Hence, within our extended objects framework, the equations 
of motion are of second order, provided that the Lagrangian $L(g_{ab}, K_{ab})$ takes the specific 
form.
\be 
L_s = \delta^{a_1 a_2 \cdots a_s}_{b_1 b_2 \cdots b_s}
K^{b_1}{}_{a_1} K^{b_2}{}_{a_2} \cdots K^{b_s}{}_{a_a},
\,\,\, (s=0,1,2, \ldots p),
\label{lags}
\ee
where we have absorbed a numerical factor for short in the 
notation. 

By taking the derivative of~(\ref{lags}) to find the 
matrix $L^{ab}_s$, and using the fact that the gKd is 
skew-symmetric under interchange of any two of the indices, 
we obtain $L^{ab} \longrightarrow L^{ab}_s = s \,g^{ac}
\delta^{ba_2 a_3 
\cdots a_s}_{c b_2 b_3 \cdots b_s} K^{b_2}{}_{a_2} 
K^{b_3}{}_{a_3} \cdots K^{b_s}{}_{a_s}$. This simple 
computation motivates the introduction of the structure
\be 
J^a_{(s)b} := \delta^{a a_1 a_2 \cdots a_s}_{b b_1 b_2 
\cdots b_s} K^{b_1}{}_{a_2} K^{b_2}{}_{a_2} \cdots
K^{b_s}{}_{a_s}.
\label{Jab}
\ee
We assert that the tensors $J^{ab}_{(s)}$ are symmetric 
and divergence-free because $\na J^{ab}_{(s)} = 0$ holds 
when the ambient background spacetime is Minkowski. This 
fact is proved by using the Codazzi–Mainardi integrability 
condition in a flat background spacetime, $\na K_{bc}
= \nb K_{ac}$,~\cite{Spivak1970,defo1995}. Technically, 
$J^{ab}_{(r)}$, ($r<s$), is related to the cofactor of 
$K_{ab}$ in the expression of the determinant~(\ref{lags}). 
It is worth noting that the $J^{ab}_{(s)}$ satisfy the 
identity
\be 
J^{ab}_{(s)} = L_s g^{ab} - s K^a{}_c J^{bc}_{(s-1)},
\label{id7}
\ee
in fulfilment of expression~(\ref{id6}).

The following is a brief list of the Lagrangians~(\ref{lags}) 
and the conserved tensors~(\ref{Jab})
\beq
L_0 &=& 1,
\\
L_1 &=& K,
\label{L1}
\\
L_2 &=& K^2 - K^a{}_b K^b{}_a = \R,
\label{L2}
\\
L_3 &=& K^3 - 3K K^a{}_b K^b{}_a + 2 K^a{}_b K^b{}_c
K^c{}_a,
\label{L3}
\\
L_4 &=& K^4 - 6 K^2 K^a{}_b K^b{}_a + 8 K K^a{}_b K^b{}_c
K^c{}_a 
\n
\\
&+& 3 (K^a{}_b K^b{}_a)^2 - 6 K^a{}_b K^b{}_c K^c{}_d 
K^d{}_a,
\n
\\
&=& \R^2 - 4 \R_{ab} \R^{ab} + \R_{abcd}\R^{abcd}, 
\label{L4}
\eeq
and
\beq 
J^{ab}_{(0)} &=& - 2 G^{ab}_{(0)} = g^{ab},
\label{J0}
\\
J^{ab}_{(1)} &=& K g^{ab} - K^{ab},
\label{J1}
\\
J^{ab}_{(2)} &=& - 2 G^{ab}_{(1)} = \R g^{ab} - 2 \R^{ab}
\label{J2}
\\
J^{ab}_{(3)} &=& L_3 g^{ab} - 3 \R K^{ab} + 6 K K^a{}_c K^{bc}
- 6 K^a{}_c K^b{}_d K^{cd} 
\label{J3}
\eeq
respectively, where $K= g^{ab}K_{ab}$, and have used the 
original notation for the Lovelock tensor in pure 
gravity~\cite{Lovelock1971, Rojas2013}, where Einstein 
tensor $G^{ab}_{(1)} = G^{ab}$ is included, in addition of 
the use of the contracted Gauss-Codazzi integrability condition 
$\R_{ab} = K K_{ab} - K_{ac} K^c{}_b$~\cite{Spivak1970,defo1995}.

Some remarks are in order. The conserved stress tensor
takes a concise form, expressed in terms of conserved tensors,
$f_{(s)}^{a\,\mu} = J^{ab}_{(s)} %+ T^{ab}_{\text{\tiny m}})
\,e^\mu{}_b$, and is entirely tangential to $m$. The 
identity~(\ref{id7}) is consistent, since for $s=n$ the 
conditions for applying the well-established Cayley-Hamilton 
theorem are met, which require that for such a value $A_{(n+1)} 
=0$,~\cite{Lovelock1989,Tapia2007}. Regarding this last point, 
only the first $n$ discriminants of $K^a{}_b$, $A_{(s)}$, 
with $s\leq n$, are non-trivial and linearly independent, 
and non-zero. Viewed in this way, by summing all the terms 
$L_s$, each scaled by a dimensionally consistent constant 
$\alpha_s$, we arrive at a robust framework analogous to the 
Lovelock theory of gravitation~\cite{Lovelock1971}. 
Clearly, 
\be 
\SLB[X^\mu] : = \int_m d^{p+1} x\,\sqrt{-g} \sum_{s=0}^{p}
\alpha_s L_s (g_{ab}, K_{ab}),
\label{LBT}
\ee
with $L_s$ given by~(\ref{lags}), also yields second-order
in derivatives equations of motion. This framework has been 
discussed in some contributions,~\cite{Rojas2013,Rojas2016,Rojas2019,Rojas2025}, 
and has been referred to as \textit{Lovelock type brane 
gravity}. In view of this, both $L_s$ and $J^{ab}_{(s)}$ 
were termed \textit{Lovelock-type brane invariants} (LBI) and 
\textit{Lovelock-type brane tensors} (LBT), respectively, 
since they represent extensions for extended objects of 
arbitrary dimension of the original geometric structures 
of the Lovelock theory,~\cite{Lovelock1971}. 
Hereafter, Lovelock type brane gravity theory will be 
considered the broadest geometrical framework, in terms of 
the fundamental forms, leading to second-order equations 
of motion where $\LLB := \sum_{s=0}^{p} \alpha_s L_s (g_{ab}, 
K_{ab})$. This leads directly to the results $( \partial 
\LLB /\partial K_{ab} ) = \sum_{s=1}^p \alpha_s \,s\, 
J^{ab}_{(s-1)}$ and $\fLB^{a\,\mu} = \sum_{s=0}^p \alpha_s
\, J^{ab}_{(s)}\,e^\mu{}_b$. 

Allowing for matter living on the brane, from~(\ref{famu2})
we have
\be 
f^{a\,\mu} = \left( \sum_{s=0}^p \alpha_s\, J^{ab}_{(s)} + 
\Tm^{ab} \right)\pb X^\mu,
\label{famu-m}
\ee
in such a way that variation~(\ref{var3}) reads
\be 
\delta S_T = - \int_m \pa \left[ \sqrt{-g} \left( 
\sum_{s=0}^p \alpha_s J^{ab}_{(s)} + \Tm^{ab}
\right)\pb X \right] \cdot \delta X
\label{var4}
\ee
with the boundary term disregarded. This yields the equations 
of motion 
\be 
\pa \left[ \sqrt{-g} \left( \sum_{s=0}^p \alpha_s J^{ab}_{(s)} 
+ \Tm^{ab} \right)\pb X^\mu \right] = 0.
\label{eom5a}
\ee
The RT model is built-in into our theory. Indeed, for that 
particular case, considering $s=2$ and the expressions~(\ref{L2}) and~(\ref{J2}) we have $J^{ab}_{(2)} = - 2 G^{ab}$, and 
choosing that $\alpha_2 = 1/2\kappa$ with $\kappa$ being a 
constant, we have that~(\ref{eom5a})
become
\be 
\pa \left[ \sqrt{-g} \left( G^{ab} - \kappa\,\Tm^{ab} 
\right)\pb X^\mu \right] = 0,
\label{eomRT}
\ee
which is clearly in accord with the well-known equations 
of motion for the GBG theory, [see~(\ref{eom0})],~\cite{RT1975,Davidson2003}.

To conclude this section, we must note that this brane 
gravity theory, must be affine in the accelerations which 
indicates that boundary terms are inherently present in each 
model,~\cite{Rojas2016b}.

%%%%%%%%%%%%%%%%%%%%%%%%%%%%%%%%%%%%%%%%%%%%%%%%%%%%%%%%%%%%%%%%%%%
\section{Dark matter and mimetic gravity effects are inherent to LBG}
\label{sec4}
%%%%%%%%%%%%%%%%%%%%%%%%%%%%%%%%%%%%%%%%%%%%%%%%%%%%%%%%%%%%%%%%%

The conservation law $\na f^{a\,\mu} =0$, even in the presence
of matter, provides both the equations of motion and the 
geometric identities that the field variables must satisfy. 
In this light, let $f^{a\,\mu }$ be subject to a particular 
shift
\be 
f^{a\,\mu} \quad \longrightarrow \quad f^{'a\,\mu} = f^{a\,\mu} 
+ \T^{a\,\mu},
\label{famu3}
\ee
where $\tau^{a\,\mu}$ is assumed to be a differentiable
vector depending on the world volume geometry. To avoid
affecting the dynamics of the brane, the condition 
$\na f^{'a\,\mu} = 0$ must hold, which impose the following restriction on $\tau^{a\,\mu}$ 
\be 
\na \T^{a\,\mu} = 0,
\label{cond1}
\ee
Therefore, the eom~(\ref{eom5a}) remains unaltered. 
Given that $\T^{a\,\mu}$ is a world volume vector, this 
allows us to decompose it into its tangential and normal 
parts,
\be 
\T^{a\,\mu} = \T^{ab}\,e^\mu{}_b + \T^a\,n^\mu.
\label{tauamu}
\ee
By plugging~(\ref{famu-m}) and~(\ref{tauamu}) into the
condition $\na f^{'\,a\mu} = 0$, bearing in mind~(\ref{cond1}) 
and the GW equations, they lead to the normal and tangential projections of $\na f^{'a\,\mu} = 0$ being given by
\be 
\begin{aligned}
- \left( \sum_{s=0}^p \alpha_s\,J^{ab}_{(s)} + 
T^{ab}_{\text{\tiny m}} \right) K_{ab} + \na \T^a - 
\T^{ab} K_{ab} &= 0,
\\
\na  \left( \sum_{s=0}^p \alpha_s\,J^{ab}_{(s)} + 
T^{ab}_{\text{\tiny m}} \right) + \na \T^{ab} + \T^a\,K_a{}^b 
&= 0,
\end{aligned}
\n
\ee
respectively. By not altering the brane dynamics and 
invoking the conservation of $J^{ab}_{(s)}$ and 
$T^{ab}_{\text{\tiny m}}$, these expressions lead to
\be 
\begin{aligned}
\na \T^a - \T^{ab} K_{ab} &= 0,
\\
\na \T^{ab} + \T^a\,K_a{}^b &= 0.
\end{aligned}
\ee
Lastly, achieving our aim requires setting $\T^a = 0$ 
to avoid resulting in a third-order equation of motion.
This leads to 
\be 
\T^{a\,\mu} = \T^{ab}\,\pb X^\mu.
\label{cond2}
\ee
Therefore, to avoid altering the extended object dynamics 
or modifying the order of the eom, the 
additional unfamiliar conserved current $\mathcal{T}^{a\,\mu}$ 
must meet the condition
\be 
\begin{aligned}
\T^{ab} K_{ab} &= 0,
\\
\na \T^{ab} &= 0.
\end{aligned}
\label{cond3}
\ee
Substituting~(\ref{cond2}) into condition~(\ref{cond1}) yields
\be 
\na ( \T^{ab}\,\pb X^\mu ) =0.
\label{cond4}
\ee 

On physical grounds, in view of~(\ref{cond3}), from the 
condition $\na f^{'a\,\mu} = (1/\sqrt{-g}) \pa \left[ 
\sqrt{-g}(f^{a\,\mu} + \T^{ab})\pb X^\mu \right] = 0$, and 
the expression~(\ref{cond2}), the eom can be written as follows
\be 
\pa \left[ \sqrt{-g} \left( \sum_{s=0}^p \alpha_s\,J^{ab}_{(s)}
+ T^{ab}_{\text{\tiny m}} + \T^{ab} \right)\pb X^\mu  
\right] = 0.
\label{eom5}
\ee
As before, for the single specific value $s=2$ with 
$\alpha_2 = 1/ \kappa$, which is the RT model case, this 
expression leads to recognizing that the structure
\be 
\R^{ab} - \frac{1}{2}\R\,g^{ab} - \kappa\, T^{ab}_{\text{\tiny m}}
= \kappa\,\T^{ab},
\label{cond5}
\ee
is conserved, complemented by the conditions~(\ref{cond3}), 
which is clearly in accord what is stated in~\cite{Davidson2003}.

Regarding the second equality in~(\ref{cond3}), this resembles 
the conservation law of an energy-momentum tensor corresponding 
to a certain fictitious matter, unrelated to the ordinary one 
included in $T^{ab}_{\text{\tiny m}}$, so we could likewise 
use the term ``dark" (with a slight abuse of language) 
to differentiate it from ordinary matter. 

By recasting the GBG theory as a mimetic gravity 
theory~\cite{Paston2018}, the embedding functions $X^\mu$, 
which act as the independent variables, serve as the auxiliary 
variables $\tilde{g}_{\mu\nu}$ and $\lambda$ of mimetic gravity~\cite{Mukhanov2013}. With that in mind, by varying 
the GBG action functional with respect to $X^\mu$, 
including matter, the field equations result in~(\ref{eom0}), 
which, by assuming the existence of an internal extra current with
components $\T^{ab}$, can be rewritten as~(\ref{cond4}).
In this light, the pair of equations consisting 
of~(\ref{FSff}) and~(\ref{cond4}),
\be 
\begin{aligned}
\pa X \cdot \pb X &= g_{ab},
\\
\na (\T^{ab}\,\pb X^\mu ) &= 0,
\end{aligned}
\label{eomDM}
\ee
are interpreted as the equations of motion that
describe this fictional matter, also called \textit{embedding 
matter}, described by the fields $X^\mu$ and $\T^{ab}$,~\cite{Paston2018}. A direct comparison reveals that 
Lovelock-type brane gravity can be recast as an embedded 
mimetic gravity theory, owing to its engaging geometric 
properties.

The extended objects framework also provides a plausible 
physical interpretation for the current $\T^{a\,\mu}$. 
Indeed, expression~(\ref{cond1}) itself suggests a 
mechanical argument that explains the nature of this 
current, by exploiting the translational invariance of 
the system, stemming from the Noether theorem,~\cite{Capo2000}. 
We begin by constructing the quantity $(\na \T^a) \cdot 
\delta X$ where $\delta X^\mu$ denotes arbitrary world volume deformations. Then, we integrate it over $m$,
\be 
\int_m \sqrt{-g} \,(\na \T^a)\cdot \delta X 
= a\cdot \int_m \pa (\sqrt{-g}\,\T^a).
\n
\ee
Here, we have considered the translation $\delta X^\mu 
= a^\mu$ of the world volume, where $a^\mu$ is a constant 
space time vector. As a consequence of the Noether theorem, 
a translational symmetry implies the conservation of the total momentum leaving unchanged the dynamics of the brane,~\cite{Capo2000}. In keeping with this symmetry, using Stokes' 
theorem on $m$ 
\be
a \cdot \int_m d^{p+1}x\,\sqrt{-g}\,\na \T^a =
a \cdot  \int_{\partial m} d^pu\sqrt{h}\,\eta_a 
\T^a = 0.
\n
\ee
where $\eta^a$ is the time-like unit normal vector pointing 
outwards to the boundary $\partial m$, such that 
$g_{ab} \eta^a \eta^b = -1$. It is built to be tangent to 
$m$, and $h := \det (h_{AB})$ with $h_{AB}$ being the metric 
induced on $\partial m$.

Grounded in the theory of elasticity,~\cite{Landau1986}, and 
given that $\T^{a\,\mu}$ directly accompanies the conserved 
stress tensor $f^{a\,\mu}$ in the definition~(\ref{famu3}), 
then the space-time vector $ \eta_a \T^{a\,\mu}dS$ is identified as 
the force per unit area $dS$ transmitted through the boundary 
element $dS = d^p u\,\sqrt{h}$ due to the action of the current $\T^{a\,\mu}$ set up within the volume~\cite{Guven2017}. 
In this sense, 
\be 
a \cdot \int_m d^{p+1}x\,\sqrt{-g}\,\na \T^a = a\cdot F = 0,
\n
\ee
where
\be 
F^\mu := \int_{\partial m} d^p u\,\sqrt{h}\,\eta_a \T^{a\,\mu}.
\label{force}
\ee
This boundary integral is identified as the total force 
exerted by internal stresses (sources inside $m$) on the 
boundary world volume $\partial m$. What happens internally 
is beside the point since it has no dynamical consequences.  
More specifically, $\eta_a \T^{a\,\mu}$ is the force per unit
area $dS$ acting on the boundary $\partial m$, due to the 
action of surface stresses~\cite{Guven2017}. Clearly, the 
net vector sum of internal forces is zero. In this sense, 
$\T^{a\,\mu}$ must be interpreted as an internal stress 
tensor, leaving unchanged the dynamics of the extended 
object. Nevertheless, no certainty that such boundary stress 
has a physical basis, but rather that it may be purely geometric.
This result is reminiscent of the fact that, in non-relativistic 
mechanics, the sum of all internal forces vanishes, resulting 
in no effect on the motion of the center of mass, and the 
equations of motion unaltered. If that is the case, the center 
of mass acts as a point particle, accelerating or moving 
at a constant rate solely in response to the net external 
forces or stresses. This conclusion was preliminarily outlined 
within a Hamiltonian framework in~\cite{Davidson1998,Davidson2003}. 
%On this 
%basis, the current $\T^{a\,\mu}$ turns out to be a 
%non-geometrical characteristic defined on the world volume, 
%involving only internal forces acting on it, and leaving its 
%overall dynamics unaffected.

%%%%%%%%%%%%%%%%%%%%%%%%%%%%%%%%%%%%%%%%%%
\section{Actions classically equivalent}
\label{sec5}
%%%%%%%%%%%%%%%%%%%%%%%%%%%%%%%%%%%%%%%%%%

As already pointed out, $\T^{a\,\mu}$ and $\T^{ab}$, related 
to internal stresses, do not affect the brane dynamics as 
their presence leaves the equations of motion unchanged. One 
might consider them as gauge-type terms. Their existence serves 
to complete the geometrical framework for describing extended objects,~(\ref{eom5}), with possible applications to depict accelerated systems. The question naturally arises as to whether 
they can be included as parts of an extended action, through 
Lagrange multipliers, resulting in a complete variational setup. 
Paston~\cite{Paston2018,Paston2020-a,Sheykin2020} has advanced 
significantly in re-engineering GBG, aiming to incorporate 
such dark matter into extended actions. By way of example, we 
will set forth two extended actions that harmonize with our development, and also yield the equation of motion for such fictitious matter.

Generally, an extended action yielding the equation of motion 
for this dark matter would have the form
\be 
S_T[ X^\mu, \lambda] = S_{\text{\tiny LBG}} + S_{\text{\tiny m}}
+ S_{1,2}[X^\mu,\lambda],
\label{actionT}
\ee
where $\lambda$ is an appropriate Lagrange multiplier and 
$S_{1,2}$ offers two choices to achieve our aim.

In the most elementary way, consider the action
\be 
S_1[X^\mu, \T^{ab}] = \frac{1}{2} \int_m \sqrt{-g}\, \T^{ab} 
\left( \pa X \cdot \pb X - g_{ab} \right)
\label{S1}
\ee
where $\T^{ab}$ plays the role of a Lagrange multiplier~\cite{Golovnev2014}. Variation of~(\ref{actionT}) with respect 
to $\T^{ab}$ immediately leads to the form of the induced metric~(\ref{FSff}),
\be 
g_{ab} = \pa X \cdot \pb X. 
\ee
Likewise, variation with respect to $X^\mu$, bearing in 
mind~(\ref{var4}), results in
\beq 
\delta S_T &=& 
- \int_m \pa \left[ \sqrt{-g} \left( 
\sum_{s=0}^p \alpha_s J^{ab}_{(s)} + \Tm^{ab}
\right) \pb X \right] \cdot \delta X
\n
\\
&+& \int_m \sqrt{-g}\,\T^{ab} \left( \pb X \cdot \pa \delta X
\right).
\n
\eeq
Integrating by parts, and neglecting a boundary term, we obtain
{\small 
\be
\delta S_T = - \int_m \pa \left[ \sqrt{-g} 
\left( 
\sum_{s=0}^p \alpha_s J^{ab}_{(s)} + \Tm^{ab} + \T^{ab}
\right) 
\pb X \right] \cdot \delta X.
\n
\ee}
By enforcing the equations of motion on the system~(\ref{eom5}), 
the following must be fulfilled
\be 
\na (\T^{ab}\,\pb X^\mu ) = 0, 
\n
\ee
thus reproducing the conservation law~(\ref{cond4}) for the 
dark current $\tau^{a\,\mu}$.

An engaging action, in line with the spirit of bimetric 
gravity~\cite{Paston2018,Hassan2011,Golovnev2017}, is provided by
\be 
S_2 [X^\mu, \T^{a\,\mu}] = \int_m \sqrt{-g} \left( \T^a 
\cdot \pa X - \text{Tr} \sqrt{g_{ac} \T^c \cdot \T^b} 
\right),
\label{S2}
\ee
where $\sqrt{•}$ denotes the square root of a matrix with 
indices $a$ and $b$, from which its trace, $\text{Tr}$, 
is then taken. It is not yet clear here that $\T^{a\,\mu}$ 
is a Lagrange multiplier, but this fact is hidden in the 
square root matrix, as will become evident below. In fact,
the constraint into the integral is nothing but the expression
of the dark current~(\ref{cond2}) written entirely
in terms of $X^\mu$ and $\T^{a\,\mu}$ itself.

To make the development plain, it is convenient to introduce the 
notation for the square root matrix
\be 
\B_a{}^b := \sqrt{g_{ac}\T^c \cdot \T^b},
\label{Bmatrix}
\ee
whereof the action~(\ref{S2}) is written in a more tractable form
\be 
S_2 [X^\mu, \T^{a\,\mu}] = \int_m \sqrt{-g} \left( \T^a 
\cdot \pa X - \B_a{}^a \right).
\label{S3}
\ee
In a like manner, for variational purposes, it is worth 
introducing the matrix
\be 
\A_a{}^b := g_{ac} \T^c \T^b - \delta_a{}^b,
\label{Amatrix}
\ee
which is a particular form of a disformal transformation.
It follows from~(\ref{Bmatrix}) and~(\ref{Amatrix}) that
\be 
\B_a{}^c \B_c{}^b = \delta_a{}^b + \A_a{}^b.
\label{id8}
\ee
Accordingly, $\B_a{}^b$ acquires the form
\be 
\label{id9}
\B_a{}^b = \sqrt{\delta_a{}^b + \A_a{}^b}.
\ee
In view of the properties of these matrices, along with their 
variations developed in appendix~\ref{app1}, we now turn to 
the variations of~(\ref{S3}). 

Given that $X^\mu$ appears in $S_2$ only within the first term, 
it follows immediately that
\be 
\delta S_2 = - \int_m \sqrt{-g}\, \na \T^a \cdot \delta X,
\ee
where we have neglected a surface term. Following the same 
discussion as in action $S_1$, %under Hamilton principle 
the conservation of dark current,~(\ref{cond1}), 
is reproduced.

Varying $S_2$ with respect to $\T^{a\,\mu}$ yields 
\be
\delta S_2 
= \int_m \sqrt{-g} \left[ \pa X \cdot \delta \T^a
- \frac{1}{2} \Bi_{bc} \delta ( \T^c \cdot \T^b - g^{cb}) 
 \right]
\n
\ee 
where we have used the variation~(\ref{id10}) and the 
definition~(\ref{Amatrix}). Using the symmetry of the
inverse matrix $\Bi$, as discussed below~(\ref{id14}), 
we arrive at
\be 
\delta S_2 =  \int_m \sqrt{-g} \left[ \pa X \cdot \delta \T^a
- \Bi_{ab} \T^b \cdot \delta \T^a  \right].
\n
\ee
The corresponding field equation is given by
\be 
\pa X^\mu = \Bi_{ab} \T^{b\,\mu}.
\label{id12}
\ee
This result reveals two key geometric structures. On the one 
hand, by dotting~(\ref{id12}) with the tangent vector 
$\pb X^\mu$ we get
\beq 
\pa X \cdot \pb X &=& \Bi_{ac} \T^c \cdot \T^d \Bi_{bd},
\n
\\
&=& \Bi_{ac} (g^{cd} + \A^{cd}) \Bi_{db},
\n
\eeq
where we have used the identity~(\ref{id11}) and the symmetry
of $\Bi$. Inserting relationship~(\ref{id8}) yields
\be 
\pa X \cdot \pb X = g_{ac} \B^{cd}\Bi_{db} = g_{ab},
\n
\ee
thus obtaining the induced metric, as expected. Likewise,
by dotting~(\ref{id12}) with the current $\T^{a\,\mu}$
we have 
\be 
\T^a \cdot \pa X = \T^a \cdot \left[ \Bi_{ab} \T^b \right]
= \Bi_{ab} (g^{ba} + \A^{ba}),
\n
\ee
where we used~(\ref{id11}). With the aid of the 
identity~(\ref{id8}) the above equation can be written in the form 
$\T^a \cdot \pa X = \Bi_a{}^b \B_b{}^c \B_c{}^a = \B_a{}^a$,
which leads to
\be
\T^a \cdot \pa X - \B_a{}^a = 0.
\label{id13}
\ee
That is, we have replicated the constraint that appears 
in the action functional $S_2$, thereby revealing that the 
dark current $\T^{a\,\mu}$ serves the role of a Lagrange 
multiplier

To render a full account of what arises from $S_2$, we can 
further derive the associated energy-momentum tensor. Under the constraint~(\ref{id13}), the non-vanishing part of the variation 
is the trace $\B_a{}^a$. Hence, from~(\ref{id10}) we have
\be 
\delta S_2 =
\int_m \sqrt{-g}\left[ -\frac{1}{2} \Bi_a{}^b
\delta (g_{bc} \T^c \cdot \T^a - \delta_b{}^a)\right],
\n
\ee
where relation~(\ref{Amatrix}) has been used. Continuing
we have
\beq 
\delta S_2 &=& \int_m \sqrt{-g} \left[ \frac{1}{2} \Bi_a{}^b
(\delta_d{}^a + \A_d{}^a)g_{bc} \delta g^{cd}\right],
\n
\\
&=&
\int_m \frac{\sqrt{-g}}{2} \B_{ab}\,\delta g^{ab},
\n
\eeq
where, in the last step, we have made use of the 
identity~(\ref{id8}) again, so that $\delta S_2 / \delta g^{ab}
= (\sqrt{-g}/2) \B_{ab}$. According to the definition for
the energy momentum tensor, we find
\be 
\T_{ab} = \B_{ab}.
\ee
We validate our result through the contraction of this 
expression with the tangent vector, $\T^{ab} \pb X^\mu = 
\B^{ab} \pb X^\mu = \B^{ab} \Bi_{bc} \T^{c\,\mu} = \T^{a\,\mu}$,
where we have considered the equation of motion~(\ref{id12}).
This reproduce the explicit form of this conserved current.

%In the same parlance, we would like to express the eom~(\ref{id12}) 
%entirely in terms of $\T^{a\,\mu}$. Taking
%the difference of the derivatives of~(\ref{id12}),
%\be 
%\nc \pa X^\mu - \na \pc X^\mu = \nc \left[ \Bi_{ab} \T^{b\mu}
%\right] - \na \left[ \Bi_{cb} \T^{b\mu} \right]
%\n
%\ee
%By considering the GW equations and the symmetry of the
%extrinsic curvature, we find
%\be 
%\na \left[ \Bi_{bc} \T^{c\,\mu} \right] - \nb \left[ \Bi_{ac}
%\T^{c\,\mu} \right] = 0,
%\ee
%thus excluding the variables $X^\mu$ since they are field variables 
%specific to the action~(\ref{action}).

%%%%%%%%%%%%%%%%%%%%%%%%%%%%%%%%%%%%%%%%%%%%%%%%%%%%%%%%%%%%
\section{Conclusions}
\label{sec6}
%%%%%%%%%%%%%%%%%%%%%%%%%%%%%%%%%%%%%%%%%%%%%%%%%%%%%%%%%%%%%

In this paper, we have formally derived the Lovelock-type 
brane gravity from a matrix analysis of the equations 
of motion that arise from general models that depend on the 
fundamental forms, structures useful for describing branes. 
The LBG equations of motion, which are second-order in the 
derivatives of $X^\mu$, constitute an extension of the GBG 
equations of motion. This is by no means coincidental, as 
it is due to the property of being affine in acceleration 
theories. By exploiting this fact, LBG can be reformulated 
as mimetic gravity, which also allows us to identify a certain 
type of fictional matter whose source is an internal covariant current, $\T^{a\,\mu}$, which is parallel to the conserved 
stress tensor. With the support of the elasticity theory, a 
possible physical explanation for this current, which 
constitutes a type of internal stress in the brane, is provided.
The current $\T^{a\,\mu}$, which lacks a clear physical or 
geometric origin, or perhaps both, must be strongly bound to 
fulfil~(\ref{cond1}). In order to explore its origin, it will be 
interesting to examine this current, within the LBG framework, 
in a highly symmetric FRW-type geometry, rewrite the associated 
Friedman-type equation, and identify the energy-momentum tensor 
$\T^{ab}$, perhaps in terms of correction extrinsic curvature
terms. This could provide further insight into its nature. 
Regarding this point, in~\cite{Paston2012,Paston2020-a,Paston2021, Paston2026}, within the framework of a non-relativistic approach, progress has been made on this origin. It is worth mentioning 
that alternative forms of the action  $S[X^\mu,\lambda]$ were 
investigated in~\cite{Paston2018}. All of that proposals lead 
to the equations of motion of the embedding matter. We would 
like to mention that these have a direct extension to LBG.

%%%%%%%%%%%%%%%%%%%%%%%%%%%%%%%%%%%%%%%%%%%%%%%%%%%%%%%%
\begin{acknowledgments}

I have benefited from conversations with Rub\'en Cordero 
and Giovany Cruz. The author acknowledges encouragement 
from ProDeP-M\'exico, CA-UV-320: \'Algebra, Geometr\'\i a 
y Gravitaci\'on. Additionally, I thank the partial support from Sistema Nacional de Investigadoras e Investigadores, M\'exico.
\end{acknowledgments}
%%%%%%%%%%%%%%%%%%%%%%%%%%%%%%%%%%%%%%%%%%%%%%%%%%%%%%%%%%%

\appendix

%%%%%%%%%%%%%%%%%%%%%%%%%%%%%%%%%%%%%%%%%%%%%%%%%%%%%%
\section{Variation of the trace of square root matrix $\B_a{}^a$}
\label{app1}
%%%%%%%%%%%%%%%%%%%%%%%%%%%%%%%%%%%%%%%%%%%%%%%%%%%%%%%%

Here, we provide the main variation behind the obtaining of 
the equations of motion from the action~(\ref{S3}), as well
as the symmetries of matrices~(\ref{Amatrix}) and~(\ref{id9}).

Varying~(\ref{id8}) yields
\be 
\delta \B_a{}^c \B_c{}^b + \B_a{}^c \delta \B_c{}^b = \delta
\A_a{}^b.
\n
\ee
Letting $\Bi_a{}^b$ denotes de inverse matrix of $\B_a{}^b$,
such that  $\Bi_a{}^c \B_c{}^b = \delta_a{}^b$.
Multiplying the previous relation on the left hand side by this 
inverse matrix we get $\Bi_a{}^c \delta \B_c{}^d \, \B_d{}^b +
\delta \B_a{}^b = \Bi_a{}^c \delta \A_c{}^b$. By taking the trace
of this expression we find
\be 
\delta \B_a{}^a = \frac{1}{2} \Bi_a{}^b \delta \A_b{}^a.
\label{id10}
\ee

Regarding the symmetries of the matrices~(\ref{Amatrix}) 
and~(\ref{id9}) we note the following. From~(\ref{Amatrix})
we readily obtain that 
\be 
\A^{ab} = \T^a \cdot \T^b - g^{ab},
\label{id11}
\ee
so that $\A^{ab}$ results symmetric. Concerning the 
matrix~(\ref{Bmatrix}), using the index-free notation,
$\B = \sqrt{\mathbb{I} + \A}$, followed of a Taylor expansion
given by 
\be 
\sqrt{\mathbb{I} + \A} = \sum_{k=0}^\infty \binom{1/2}{k}
\A^k = \mathbb{I} + \frac{1}{2} \A - \frac{1}{8} \A^2 +
\frac{1}{16} \A^3 + \cdots
\ee 
with the special binomial coefficient
defined as $\binom{r}{k} = [r(r-1)(r-2) \cdots (r-n+1)]/k!$ being
$k$ a positive integer. Further, we will assume that the root branch
is taken so that $(\mathbb{I})^{1/2} = \mathbb{I}$. Given these 
aspects, we get
\be 
\B^{ab} = g^{ba} + \frac{1}{2} \A^{ba} - \frac{1}{8} 
\A^{bc}\A_c{}^a + \cdots ,
\label{id14}
\ee
in order that $\B^{ab}$ is symmetric. Lastly, we point out 
that if the matrix~(\ref{Bmatrix}) is 
non-singular and symmetric, then the inverse matrix $\Bi$ is 
symmetric, $\Bi_{ab} = \Bi_{ba}$.

%%%%%%%%%%%%%%%%%%%%%%%%%%%%%%%%%%%%%%%%%%%%%%%%%%%%%%%%%%%%%

%\section{Appendixes}

%To start the appendixes, use the \verb+\appendix+ command.

%\nocite{*}

\bibliography{mimetic-love}% Produces the bibliography via BibTeX.

\end{document}